# Ultraviolet Photodetectors based on GaN and AlGaN/AlN Nanowire Ensembles: Effects of Planarization with Hydrogen Silsesquioxane and Nanowire Architecture

E. Akar[†,*], I. Dimkou[‡], A. Ajay[†], Martien I den Hertog[§], and E. Monroy[†]

[†] Univ. Grenoble-Alpes, CEA, Grenoble INP, IRIG, PHELIQS, 17 av. des Martyrs, 38000 Grenoble, France
[‡] Univ. Grenoble-Alpes, CEA, LETI, 38000 Grenoble, France
[§] Univ. Grenoble-Alpes, Institut Néel-CNRS, 25 av. des Martyrs, 38000 Grenoble, France

*Corresponding author: elcin.akar@cea.fr

OrcID:
Elçin Akar: 0009-0002-4221-3357
Ioanna Dimkou: 0000-0002-6865-3820
Akhil Ajay: 0000-0001-5738-5093
Martien I den Hertog: 0000-0003-0781-9249
Eva Monroy: 0000-0001-5481-3267

## Abstract

The interest in nanowire photodetectors stems from their potential to improve the performance of a variety of devices, including solar cells, cameras, sensors, and communication systems. Implementing devices based on nanowire ensembles requires a planarization process which must be conceived to preserve the advantages of the nanowire geometry. This is particularly challenging in the ultraviolet (UV) range, where spin coating with hydrogen silsesquioxane (HSQ) appears as an interesting approach in terms of transmittance and refractive index. Here, we report a comprehensive study on UV photodetectors based on GaN or AlGaN/AlN nanowire ensembles encapsulated in HSQ. We show that this material is efficient for passivating the nanowire surface, it introduces a compressive strain in the nanowires and preserves their radiative efficiency. We discuss the final performance of planarized UV photodetectors based on three kinds of nanowire ensembles: (i) non-intentionally-doped (nid) GaN nanowires, (ii) Ge-doped GaN nanowires, and (iii) nid GaN nanowires terminated with an AlGaN/AlN superlattice. The incorporation of the superlattice allows tuning the spectral response with bias, which can enhance the carrier collection from the AlGaN/AlN superlattice or from the GaN stem. In all the cases, the performance of the planarized devices remains determined by the nanowire nature, since their characteristics in terms of linearity and spectral selectivity are closer to those demonstrated in single nanowires than those of planar devices. Thus, the visible rejection is several orders of magnitude and there is no indication of persistent photocurrent, which makes all the samples suitable for UV-selective photodetection applications.

**KEYWORDS:** Photodetector, ultraviolet, GaN, nanowire, superlattice, planarization, photoconductor, photodiode

## 1. INTRODUCTION

The demand for reliable and high quality detection of ultraviolet (UV) radiation has been growing in importance owing to its use in distinct application domains such as pollution monitoring, flame detection, missile tracking, assessment of erythema/cancer hazard or space communication.[1–3] Such trend has led an advancing research on the UV photodetectors to improve the device performance with innovative ideas.[4–11] A high-performance photodetector should satisfy the "5S" requirements: high *sensitivity*, high *signal-to-noise ratio*, high spectral *selectivity*, high *speed*, and high *stability*.[12] Silicon-based detectors are widely used due to their well-developed and low-cost technology for device fabrication. However, their spectral selectivity relies on the use of filters, which are bulky and present lifetime issues.[13] With direct bandgap and the possibility to control n- and p-type doping, III-nitrides have emerged as excellent materials for the development of high-performance UV photodetectors with a spectral cutoff in the 200-400 nm range.[4,14–17] Depending on the specifications, various photodetection mechanisms can be exploited ranging from photoconductor, Schottky photodiode, p-n and p-i-n junction photodiodes, avalanche photodiode to metal-semiconductor-metal photodiodes by modifiying the device architecture.

The nanowire geometry has opened new perspectives in the field of photodetection.[14,18] It is a promising approach since such structures can be grown almost defect-free with high surface-to-volume ratio. During growth, the strain relaxation along the surface can prevent dislocation formation. Recently, the performance of single-nanowire devices has been investigated in detail.[9,19] Single-nanowire photoconductors are characterized by high photocurrent gains, which can reach $10^6$, and strong spectral contrast above and below the bandgap. A common feature in nanowire photoconductors is the fact that the photocurrent scales sublinearly with the impinging laser power, which has been shown for single GaN nanowires regardless of the presence of heterostructures.[20–24] In the case of undoped GaN nanowires, Sanford *et al.* and Spies *et al.* reported an improvement of the linearity in nanowires with small diameter (< 100 nm), which they attributed to the total depletion of the nanowires.[20,25] More recently, Cuesta *et al.*[26] studied a single GaN with an axial p-n junction. They showed that under reverse bias the effect of surface states becomes less emphasized and the responsivity increases linearly with a decay time of ~10 μs.

Apart from single nanowires, UV photodetectors based on nanowire ensembles are promising candidates for large area UV photodetection.[27,28] In this case, the implementation of the top (illuminated) contact of the device requires the planarization of nanowires, generally obtained by spin coating of a polymer. Different polymers have been employed as planarizing materials, such as polyimide (PI)[29], parylene[30], polydimethylsiloxane (PDMS)[31] and the most commonly utilized hydrogen silsesquioxane (HSQ).[32–35] One of the first studies on nanowire ensembles based on a GaN p-i-n junction was performed by de Luna Bugallo *et al.*[32]. They fabricated UV photodetectors by planarizing with HSQ and using indium tin oxide (ITO) as top contact. The device presents a visible rejection of the order of ~$10^2$ at zero bias. However, the response decreases for energies higher than 3.5 eV (≈ 354 nm) since ITO starts to absorb. Also, it shows linear response as a function of incident power at low power levels (< $6\times10^{-4}$ W/cm$^2$), however subsequently the linearity deviates as a result of trapping/releasing of photogenerated carriers. The photodetector decay time was measured in the order of ms, and it is associated with a photoconductive gain mechanism appearing both at forward and reverse bias.[36] Another research[33] demonstrated the fabrication of photodetectors based on an HSQ-planarized GaN nanowire ensembles using few-layered graphene as a transparent electrode. Similar to the previous example, the device shows linear behavior at low power and becomes sub-linear with increasing power density. The device achieves responsivity of 25 A/W under 1 V at 357 nm; unfortunately, no time response analysis was reported on this device. Furthermore, Aiello *et al.*[30] utilized parylene polymer for the planarization of InGaN/GaN disk-in-nanowire ensemble, showing photodetectors that exhibit rise and decay time response of 190 s and 454 s respectively. So far, planarization was considered a part of the fabrication process, but there is little analysis (if any) of the effect of the planarizing material on the optical properties of the nanowires and on device performance. This is a relevant question to assess to what extent the advantages of single nanowire photodetectors can be extended to devices based on arrays of nanowires.

In this study, we discuss the effect of planarization with HSQ on the optical properties of GaN nanowires, and we report a fabrication process for planarized UV photodetectors based on GaN nanowire ensembles. The process is applied to three different nanowire samples grown by plasma-assisted molecular beam epitaxy, including undoped and Ge-doped GaN nanowires, and GaN nanowires terminated with an AlGaN/AlN superlattice, to shift the spectral response to shorter wavelengths. The devices are investigated in terms of linearity, responsivity, spectral selectivity, frequency dependence, and time response. The results demonstrate that devices based on nanowire ensembles present characteristics that are closer to those of single nanowires rather than to planar structures. In particular, they maintain excellent spectral selectivity and faster response time than planar photoconductors.

## 2. EXPERIMENTAL SECTION

**2.1. Samples under Study.** In this work, we discuss the photodetector fabrication process and final performance of a set of three kind of nanowire ensembles: **SGaN** consisting of 1-µm-long non-intentionally-doped GaN nanowires, **SGaNGe** consisting of 1-µm-long GaN nanowires doped with Ge, and GaN nanowires terminated with an $Al_{0.1}Ga_{0.9}N$/AlN superlattice referred to as **SSL**. A scheme of the three samples is presented in Figure 1a.

**2.2. Growth of Self-Assembled GaN Nanowires.** Self-assembled GaN nanowires were synthesized on n-type Si(111) substrates using plasma-assisted molecular beam epitaxy, following the catalyst-free technique first demonstrated by Prof. Kishino and Prof. Calleja.[37,38] The substrate was degreased using dichloromethane, acetone and methanol before baking it *in situ* at 880°C to remove the native oxide. To minimize the nanowire tilt and the residual planar growth, the process started with the deposition of a low-temperature AlN buffer layer consisting of 1.2 nm deposited at 300°C followed by 8 nm at 670°C, both grown at stoichiometric conditions (Al/N flux ratio $\Phi_{Al}/\Phi_N = 1$) and with a growth rate of 450 nm/h.[39–41] Then, the substrate temperature was increased to 810°C, and the GaN nanowire growth proceeded under N-rich conditions (Ga/N flux ratio $\Phi_{Ga}/\Phi_N = 0.25$), with a growth rate of 330 nm/h. Figure 1b illustrates the growth process of the nanowires. As reported by Consonni et al.,[42] the nanowire nucleation mechanism includes different stages (see Figure 1c). Firstly, the incoming Ga and N atoms nucleate on the AlN buffer layer. Subsequently, there is an evolution of the shape of the GaN seed, to become a nanowire, whose vertical growth is fed not only by the Ga atoms impinging on top, but also by those diffusing along the sidewalls.[42] N-type doping of the nanowires was performed by incorporating an additional Ge flux during growth,[43,44] using a Ge cell temperature that leads to [Ge] = $10^{20}$ cm$^{-3}$ in planar layers.

In the case of nanowires terminated with an AlGaN/ AlN superlattice, the growth started with the same low-temperature AlN buffer layer, followed by a 900-nm-long GaN stem, to improve the uniformity of the nanowire height in the ensemble.[45] The growth continued with the deposition of an 88-period Al$_{0.1}$Ga$_{0.9}$N/AlN (1.5 nm/ 3 nm) superlattice. The Al$_{0.1}$Ga$_{0.9}$N sections were grown under N-rich conditions ($\Phi_{Ga}/\Phi_N = 0.25$ and $\Phi_{Al} = \Phi_{Ga}/9$). The AlN sections were grown at stoichiometry. In the case of AlN, the substrate temperature is too low to promote the diffusion of atoms along the sidewalls, and only the atoms impinging directly on the wire contribute to the growth.[46] The complete superlattice was synthesized without any growth interruption.

Top-view and cross-section scanning electron microscopy (SEM) images of the three samples under study are shown as Supporting Information, Figure S1. The images were of obtained in a Zeiss Ultra 55A microscope operated at 3-5 keV with the in-lens secondary electron detector. The aforementioned growth process leads to N-polar GaN nanowires with a density around 6-8×10$^9$ cm$^{-2}$ and an average diameter of 66±12 nm, 86±16 nm, and 100±21 nm for sample SGaN, SGaNGe, and SSL respectively.

A detailed view of the superlattice in sample SSL is presented as Supporting Information, Figure S2. High angle annular dark field (HAADF) scanning transmission electron microscopy (STEM) studies were performed using a Jeol Neo Arm working at 200 kV equipped with a probe aberration corrector.

**2.3. Device Fabrication.** To planarize the nanowire ensembles, FOx-25 resist (also known as HSQ) was deposited using the spin coating technique at 5000 rpm for 1 min, which leads to a nominal thickness around 580 nm on a flat surface. Details on the effect of the spinning speed on the thickness and penetration behavior of HSQ are given as Supporting Information, Table S1 and Figure S3. Subsequently, the spin-coated sample was pre-baked at 180°C for 5 min on a hotplate, and then annealed at 600°C in N$_2$ atmosphere for 1 hour. During annealing, the Si-H bonds dissociate whereas siloxane (Si-O-Si) bonds form and HSQ transforms into SiO$_x$, mostly considered as silica (SiO$_2$) due to its network-like structure.[47]

To expose the tips of the planarized nanowires, HSQ was etched in an inductively coupled plasma - reactive ion etching (ICP-RIE) Oxford Plasmalab System100 using a $CF_4/Ar/CH_2F_2$ (20/50/5 sccm) chemistry, with a radio-frequency power of 500 W and a gas pressure in the chamber of 12 mTorr, attaining an etching rate of 35 nm/min. The back contact of the devices was formed by electron beam evaporation of Ti/Au (50 nm/150 nm) directly on the back side of the Si substrate. The front contact was defined by optical lithography and electron beam evaporation of Ti/Au (50 nm/150 nm). The last step was to deposit a semitransparent (7.5 nm) Au layer by optical lithography and electron beam evaporation on the already-defined top contact pattern.

The morphological properties of the samples were probed throughout the device fabrication process using SEM and atomic force microscopy (AFM). For AFM images, we used a Bruker Dimension Icon system equipped with TESPA-V2 tips from Bruker, and operated in the tapping mode.

**2.4. Optical Spectroscopy.** The photoluminescence (PL) setup used for this study includes a frequency-doubled solid-state $Ar^+$ laser emitting at 244 nm. The photons emitted from the sample were studied using a Jobin Yvon HR460 monochromator coupled with a liquid-nitrogen-cooled UV-enhanced charge-coupled device (CCD) camera. The PL experiments were carried out at 5 K to reduce the influence of phonons or thermally activated non-radiative processes. The samples were mounted on a cold finger of a He cryostat.

**2.5. Opto-Electrical Characterization.** Current-voltage (I-V) measurements were conducted using an Agilent 4155C semiconductor parameter analyzer connected to a probe station. The validated devices in each sample were wire bonded on the ceramic chip which was then mounted on the custom-made sample holder. To evaluate the variation of the photocurrent as a function of the impinging optical power and chopping frequency, samples were excited with an unfocused continuous-wave HeCd laser ($\lambda$= 325 nm) with a spot diameter of 1 mm. For the power dependence measurements, the laser was chopped at 90 Hz and the devices were connected in series with a power supply and a Stanford Research Systems SR 830 lock-in amplifier. Depending on the device response, the lock-in amplifier is connected through a $\times 10^6$ V/A transimpedance amplifier or used as a voltmeter. In the latter case, it measures the voltage drop in a load resistor (10 k$\Omega$ or 100 k$\Omega$) connected in series with the photodetector. The same setup was connected with a TDS2022C oscilloscope in series to characterize the time response of the devices by measuring the voltage drop in the load resistance. The spectral response of the devices was characterized exciting them with a 450 W Xe lamp coupled with a Gemini 180 Jobin-Yvon grating monochromator. All the electrical and photocurrent studies were performed at room temperature.

## 3. RESULTS AND DISCUSSION

**3.1 Device Fabrication Choices: Justification and Impact.** The device fabrication process, schematically described in Figure 2, consists of the following steps: planarization of the nanowire ensembles, etching to expose the upper part of the nanowires, back and front metallization, and deposition of a semitransparent spreading layer. As a first step, it is necessary to encapsulate the nanowires in an insulating material with good wetting properties, to reduce the risk of metal evaporation between the nanowires creating shunt paths.[48] Ideally, the planarizing material should also be able to passivate surface states.[30] Furthermore, to preserve the optical advantages of the nanowire geometry, it is important that its refractive index is lower than the refractive index of the nanowires, and it must be transparent at the operation wavelength of the device.

In this study, we used the FOx-25 resist to encapsulate the nanowires. FOx (flowable oxide), also known as HSQ (hydrogen silsesquioxane), is a negative resist mostly employed for electron-beam lithography with favorable mechanical stability, good dielectric properties and high gap-filling behavior.[32] HSQ was deposited using the spin coating technique, and annealed to form silica (see Device Fabrication). With a refractive index in the range of 1.55 to 1.4,[16] to be compared with 2.52 for GaN at 400 nm,[49] and excellent transmittance in the UV range, silica fulfills the optical requirements for this application.

Since the transparency in the UV range of the planarizing material is determining for the device performance, the optical transmittance was verified experimentally: a sapphire wafer was spin-coated with HSQ at 3000 rpm for 60 s (HSQ layer thickness = 695 nm) followed by annealing at 700°C in $N_2$ atmosphere for 1 hour. The transmission was tested using a HeCd laser with $\lambda = 325$ nm at normal incidence, using uncoated sapphire as a reference. The optical transmission losses were negligible within the error bar of the measurements. Considering the planarization of SGaN nanowires, Figure 3a shows a SEM image of the cross section and an AFM image of the top surface of SGaN after spin-coating. HSQ fills the gap between nanowires and offers a flat top surface with a root-mean-square (RMS) roughness of 0.78 nm, measured in 1×1 µm². The total HSQ thickness of the coated sample is larger than the nominal HSQ thickness on a flat substrate (see Supporting Information, Table S1), revealing that the nanowire shape and density obstructs HSQ spreading. In this regard, deviations in the coating thickness are expected from sample to sample. For instance, the tapered geometry and larger diameter of the nanowires in sample SSL result in the generation of small voids during the HSQ coating, as shown in the Supporting Information, Figure S3.

Following the planarization step, dry etching was performed to expose the tip of the nanowires. Figure 3b and 3c show cross-section SEM and AFM images of the SGaN sample after etching 150 nm and 350 nm. After 150 nm etching, the surface of the sample was still uniformly covered by HSQ, without major degradation of the surface roughness (RMS roughness = 0.95 nm, measured in 1×1 µm²). After 350 nm etching, the tips of the nanowires were exposed.

From a comparative analysis of as-grown SGaN and the specimens in Figure 3b and 3c (coated SGaN after 150 nm etching and after 350 nm etching), we gain access to some optical and mechnical effects of coating on the nanowires. The low-temperature (5 K) PL spectra of as-grown and gradually etched SGaN samples are displayed in Figure 4. The spectrum of the as-grown sample is dominated by the donor-bound exciton transition ($D^0X$) at around 357 nm.[50] The free exciton A appears as a shoulder at 356 nm. In the case of the 150-nm-etched sample, with nanowires fully embedded in HSQ, the $D^0X$ peak is blue shifted and drastically broadened. These alterations can be attributed to the inhomogeneous strain distribution in the nanowires after the planarization.[51–53] If we assume that the HSQ generates a hydrostatic pressure, a blue shift on the PL spectrum can be attributed to the fact that the strain is compressive.[53]

To assess if there is a degradation of the total light emission due to the planarization process, we calculated the integrated PL intensity in both as-grown and 150-nm-etched samples, concluding that the total intensity drops by only 33% due to the encapsulation. To investigate the origin of this drop, we need to assess also the difference in excitation. In the planarized sample, the exciation laser (244 nm wavelength) has to penetrate around 200 nm of HSQ before reaching the nanowires (see Figure 3b). Therefore, we measured the transmittance of HSQ deposited on sapphire at 244 nm. The transmission loss due to the 695-nm-thick HSQ layer was 84%, with respect to the uncoated sapphire reference, which leads to an estimation of the absorption coefficient $\alpha(\lambda = 244 \text{ nm}) = 2 \times 10^4 \text{ cm}^{-1}$. Therefore, we approximate an attenuation of the excitation laser by 32%, which explains the decrease in the PL intensity.

Let us focus now on the spectrum of the 350-nm-etched sample. The $D^0X$ transition is spectrally shifted towards the as-grown sample compared to the 150-nm-etched sample with an intermediate broadening of the spectrum. Such behavior is ascribed to the release of the strain in the nanowires that occurs when exposing the tips. In this case, the integrated PL intensity is 98% of the as-grown sample value. This validates that there is no optical degradation introduced in the active material due to the planarization process, and the compressive strain can be elastically released.

To complete the study of the PL spectra, a secondary peak associated with excitons bound to surface states (SX) is observed at 359 nm in the as-grown sample.[54] In both 150-nm and 350-nm-etched samples, this peak seems to diminish which points to successful surface passivation. Finally, a transition related to excitons bound to stacking faults (SF) is observed at 362 nm in the spectrum of the as-grown sample.[54–56] Such defects are generated in the early stages of growth of self-assembled GaN nanowires. A quenching of the SF emission after planarization is unexpected, since HSQ is presumed to not alter the structural properties of the nanowires. The fact that the SF emission is not observed in samples with HSQ is due to the fact that the stacking faults are located in the vicinity of the substrate, and the probe laser would have to penetrate deep into the structure to excite that area (at 244 nm, only 13% of light can penetrate through 1 µm of HSQ).

Once the samples were planarized and etched to expose the nanowire tips, metal pads were deposited using optical lithography and electron beam evaporation of Ti/Au. The top contact incorporates a region of 375 µm × 460 µm with a pattern consisting of 3 µm fingers with 3 µm pitch. Finally, a second lithography step is applied to deposit a 7.5-nm-thick semitransparent Au layer, to increase the number of nanowires that are contacted. A top-view image of the final devices is presented in Figure 5a. The total dimension of one device, including fingers, contact pad, and semitransparent contact, is 500 µm × 500 µm.

**3.2. Electrical Characterization in the dark.** Figure 5b displays typical current-voltage (I-V) characteristics of the samples in the dark, on a semi-logaritmic scale. Bias was applied to the top contact, whereas the backside contact was grounded. SGaN exhibits a nonlinear, asymmetric characteristic, displaying higher conduction for forward bias. This indicates that the top contact is rectifying. Indeed, Ti is expected to form a Schottky contact on n-type GaN, with a Schottky barrier experimentally measured to be in the range of 0.55-0.65 eV.[57] In the case of the bottom contact, outdiffusion of silicon from the substrate is expected to slightly n-dope the bottom part of the nanowires, favoring carrier exchange with the substrate via tunnelling processes.

The I-V characteristic of SGaNGe shows improved symmetry and significantly higher current than that of SGaN. This is explained by the fact that Ge doping increases the nanowire conductivity and decreases the depletion region width below the top Schottky contact, which favors carrier tunnelling.

Finally, the I-V curve of the SSL sample presents rectifying behavior, as expected from its assymetric nature. The presence of the AlGaN/AlN heterostructure raises the Schottky barrier height of Ti/Au contact as well as the nanowire series resistance, which explains the significantly lower current level in this sample.

**3.3. Characterization under Illumination.** Figure 6 presents the photoresponse of samples SGaN and SGaNGe. The variation of the photocurrent, $I_{ph}$, as a function of the incident power density, $P$, is displayed in Figure 6a and 6b in log-log plots, where the slope of a linear trend is indicated with a dashed line. The measurements were performed at 325 nm using a 100 kΩ load resistance. The experimental data were fitted to a power law $I_{ph} \propto P^{\beta}$. The β values extracted from each curve are given on the graphs next to their corresponding curves.

To understand the response of the devices under illumination, we must keep in mind the difference between photoconductive devices and Schottky photodiodes. In photoconductors, there are two main mechanisms involved in the increase of conductance during illumination. On the one hand, photons generate additional free carriers, which is generaly a linear process with the optical power. On the other hand, light modifies the occupancy of defect states (point defects, dislocations, surfaces), which leads to a decrease of their capture radius, i.e. an increase of the average mobility. This latter process is strongly nonlinear and can lead to a huge gain and slow recovery times.[15] On the contrary, the photocurrent response of Schottky photodiodes is generally linear, since the variation of the photocurrent is dominated by the photogeneration and drift by the built-in electric field.[15] In the case of single-nanowire devices, most authors report a sublinear behavior with the optical power,[21] even in the case of Schottky diodes.[20,25,58] This feature is assigned to the relatively high series resistance associated with the nanowire stem, which behaves as a photoconductor in series with the photodiode. The nonlinear behavior of photoconductors is enhanced in the nanowire geometry due to the large surface-to-volume ratio. In general, a depletion layer forms due to the Fermi level pinning on the sidewalls of the nanowires, and the diameter of the conductive core depends on the illumination intensity, which modulates the charge state of the surface levels at the sidewalls.[58] Linearity is recovered in the case of nanowires that are thin enough to be fully depleted as the result of surface states.[25,58] In this case, the modulation of the occupancy of the surface states does not have any significant effect in the core of the wire, and photogeneration of carriers is the dominant detection mechanism.

Looking back to Figure 6a,b, the behavior of the two samples under bias is strongly sublinear, with the undoped sample providing a significantly higher response. This is consistent with the response of planar photoconductors.[59] In contrast with the doped sample, the undoped nanowire array behaves assymetrically with bias, with a higher response under reverse bias. This confirms the Schottky nature of the SGaN top contact, which plays a role to enhance the carrier collection at zero bias and under reverse bias.

Let us remind here that a symmetric photoconductor should not present any photoresponse at zero bias. However, both samples present a response, which appears due to the device asymmetry (Ti/Au top contact directly on the nanowires, and back contact on silicon, requiring carriers to traverse the heterojunction). Interestingly, there is a major difference between the two samples at zero bias: The zero-bias response of SGaN is approximately linear. Comparing with results obtained in single nanowires, this would be consistent with the nanowires being fully depleted, which is expected in view of the nanowire diameter (average nanowire diameter of SGaN ≈ 75 nm). Hence, the photogeneration of carriers dominates the response, since the variation of the surface depletion region under illumination is negligible. To validate this assumption, we calculate the critical diameter for total depletion, which can be estimated as[58,60]

$$d_{crit} = \sqrt{\frac{16\varepsilon_{GaN}\varepsilon_0 \psi}{e^2 N_D}} \tag{1}$$

where $\varepsilon_{GaN} \approx 9$ is the dielectric constant of GaN, $\varepsilon_0$ is the permittivity of vacuum, $\psi$ is the location of the Fermi level below the conduction band at the surface, $e$ is the elementary charge and $N_D$ is the net n-type dopant density. In n-type GaN nanowires, $\psi \approx 0.6$ eV, which implies the SGaN nanowires would be depleted for $N_D < 8.5 \times 10^{17}$ cm$^2$, significantly higher than the expected residual doping. On the contrary, in SGaNGe, the critical diameter to obtain complete depletion is estimated at $d_{crit} = 6.9$ nm, much smaller than the nanowire diameter observed in the SEM images (average nanowire diameter of SGaNGe ≈ 87 nm). Therefore, the conductive section of the wires is sensitive to illumination and the response at zero bias is sublinear.

Figure 6c displays the spectral response of the SGaN and SGaNGe at 5 V and 0 V, respectively. Note that the spectral responsivity measurement is done with a Xe lamp (lower power density than HeCd laser), hence SGaN sample could not be measured at low bias due to the high noise level. However, considering the symmetry of the samples, no spectral variation with bias is expected. Both samples display a sharp cut-off around λ ≈ 364 nm (i.e., GaN bandgap). The visible response remains below the noise level. This confirms that the data presented in Figure 6a and 6b correspond to the photoresponse of GaN nanowires, and are not influenced by the silicon substrate. Furthermore, it confirms that the devices maintain the spectral characteristics observed in single nanowire photodetectors: a sharp spectral cutoff even in the photoconductor configuration.[21] Planar photoconductors present a degraded spectral contrast due to the presence of extended defects generating electronic levels within the bandgap. In contrast, the main extended defect in nanowires is the sidewall surface,[15] generally *m*-{10-10} planes which do not contain mid-gap energy levels.[21] Therefore, Figure 6c shows that the spectral response of nanowire arrays is determined by the nature of the nanowire surfaces, which is not perturbed by the planarization process.

The frequency dependence of the photoconductors was measured under a 325-nm irradiance of 10 mW/cm$^2$, with the results shown in Figure 6d. The graph shows the typical behavior observed in single GaN nanowire photodetectors[21], with the photocurrent decreasing with increasing chopping frequency, which confirms the presence of a detection mechanism with slow recovery time. Consistently, the photocurrent decays measured in the oscilloscope using a 100 kΩ load resistance are in the range of hundreds of microseconds.

Finally, Figure 7 presents the photoresponse of the SSL sample: terminated with an AlGaN/AlN superlattice. As shown in Figure 7a, the SSL sample presents a nonlinear trend which can be attributed to the photoinduced variation of the resistance of the GaN stem that plays an important role during the carrier collection process. The spectral response of SSL under different bias is given in Figure 7b. The spectrum under zero bias shows the characteristic of GaN, which supports that mostly the GaN stem contributes to the photocurrent measurements. Yet, for forward bias higher than 1 V, a shift towards shorter wavelengths is observed, which suggests that the AlGaN/AlN superlattice dominates the response of the device.

It is interesting to correlate the observed spectral response with the frequency dependence as a function of bias, displayed in Figure 7c. As it was the case for the photoconductive samples, the photocurrent decreases with increasing chopping frequency for forward bias at +3 V. However, the trend is inverted for any bias below +3 V. To understand the inversion of the trend, the photocurrent as a function of time at +3 V and -3 V is given in Figure 8a and b. At +3 V, the device response exhibits rise and decay transients that are well fitted with a biexponential function. In contrast, at -3 V, the (negative) photocurrent increases sharply when the illumination is turned on, and then decreases to a stand-by level. When light is switched off, there is a current spike in the opposite direction, and then the signal progressively decreases to zero level. These transients are also well fitted with biexponential functions.

In the structure terminated with a superlattice, the electric fields involved in the collection of photogenerated carriers are schematically reprensented in Figure 8c. There is the electric field associated with the top Schottky diode ($E_{Sch}$), which points along the [000-1] direction. On the other hand, the internal electric fields due to the difference of polarization between the GaN stem and the superlattice ($E_{P,GaN}$ and $E_{P,SL}$) point towards the negatively charged interface, i.e. the field is [000-1]-oriented in the GaN stem and [0001]-oriented in the superlattice. To see what are the dominating fields and understand the behavior observed under positive/negative bias, one-dimensional calculations of the SSL band structure along the growth axis were performed using the Nextnano$^3$ software.[61,62] The results of the simulation can be seen in Figure 8d and e, under forward and reverse bias, respectively. The black band diagram corresponds to the potential profile of the nanowire structure under zero bias (the quantum wells were smoothed out in this reference profile). At zero bias, a polarization-induced negative charge sheet is formed at the interface between the GaN stem and the superlattice, hence the bands present a triangular shaped potential. Such polarization and the associated electric fields create a wide depletion region, so that the superlattice is almost fully depleted. Under positive bias, the depletion region shrinks towards the GaN/SL interface, the bands flatten along the GaN stem, and photogenerated carriers are collected mainly from the superlattice. Under negative bias, the space charge region extends both along the GaN stem and the superlattice, the bands flatten along the superlattice, and the carrier collection takes place predominantly from the GaN stem. The results of the band simulations are in good agreement with the spectral responsivity of SSL sample under different bias, showing a blue shift of the response under forward bias attesting absorption in the superlattice. Such spectral variation of the response with bias is in agreement with previous reports in single nanowire photodetectors incorporating GaN/AlN superlattices.[63,64]

The transient behavior represented in Figure 8a and b is a manifestation of the two opposed electric fields, $E_{P,GaN}$ and $E_{P,SL}$, in areas with different carrier dynamics. Under forward bias, the electric field associated with the Schottky contact, $E_{Sch}$, is negligible, and the carriers are collected from the superlattice, with photocurrent decay/rise times of $\approx 280$ µs (measured on a 100 kΩ load resistance). Under reverse bias, the contribution of $E_{Sch}$ becomes more pronounced and generates a nonuniform field distribution which may lead to a sudden increase/decrease of photocurrent during switch on/off.[65] The observed photocurrent overshoot is due to the asymmetric carrier dynamics arising from this unbalanced opposing electric fields with $E_{P,GaN}$ and $E_{Sch}$ pointing towards the Schottky contact (positive photocurrent) and $E_{P,SL}$ pointing towards the GaN stem (negative photocurrent). SSL presents fast rise time ($< 100$ µs) due to the fast reaction of the GaN stem, followed by a slow relaxation (decay time of $\approx 280$ µs), due to the compensating reaction at the SL. This explanation is consistent with the observation of a higher photoresponse when the photocurrent generated in the SL is dominant (+3 V bias in Figure 7a). The SSL sample shows increased photocurrent levels with increasing chopping frequencies for any applied bias below 3 V, Figure 7c. This can be attributed to the current spike at the onset of illumination for reverse bias (Figure 8b) increasingly influencing the average photocurrent output during the increasingly short illumination period.

Overall, all the devices present a visible rejection higher than two orders of magnitude due to the absence of energy levels in the bandgap at the nanowire sidewalls.[66] The rise and decay time response of all samples are in the range of few hundreds µs with the fastest response of $\approx 100$ µs for the SSL sample. The measured device operation speeds are faster than previously reported in photodetectors based on ensembles of GaN nanowires,[31,32] and are slower than those of ZnO single nanowires.[67] Moreover, the responsivity of the samples can be estimated as $R_{ph} = I_{ph}/(PA_{dev})$, where $A_{dev}$ is the optical area of the device. For an irradiance around 10 mW/cm², the responsivity of the samples is around 0.1-1 mA/W. These results are orders of magnitude lower than some previous reports on photodetectors based on GaN nanowire ensembles.[32,33] However, in these examples, the devices exhibit photoconductive gain, consistent with the observation of few-ms time constants. Compared with the literature, the relatively low responsivity values of our samples can be related to the semitransparent Au layer, which is not sufficient to collect the photocurrent from all the nanowires located between the fingers. Another reason could be the contact fingers themselves concealing a big part of the illuminated area. Different approaches to improve the responsivity can be applied, namely, the thickness of the Au layer can be increased to collect the signal from nanowires more efficiently. Another way could be the modification of the contact design to eliminate the effect of fingers on the nanowire illumination. Finally, using different contact materials such as ITO[30,36], graphene[28,33] or applying a silver nanowire dispersion to enhance the wires contacted in parallel[31] could be interesting to further enhance the device performance.

## ■ CONCLUSIONS

In this paper, we have assessed the effect of the planarization on the optical characteristics of GaN self-assembled nanowire ensembles grown by catalyst-free plasma-assisted molecular beam epitaxy, with the target of fabricating large-area UV photodetectors. Considering the wetting behavior, refractive index, and transparency in the UV range, we decided to perform our experiments with HSQ, which can efficiently fill the space between nanowires. A detailed photoluminescence analysis showed that HSQ planarization is successful for the surface passivation of nanowires and preserves the radiative efficiency, in spite of imposing a compressive mechanical strain in the structure. Using these results, we have established a photodetector fabrication protocol and we demonstrate a comparative study of three nanowire samples, consisting of non-intentionally-doped GaN nanowires (SGaN), Ge-doped nanowires (SGaNGe) and GaN nanowires terminated with an AlGaN/AlN superlattice (SSL). All three samples show a sublinear trend as a function of the impinging optical power, the only exception being the SGaN sample at 0 V, due to its nanowire diameter being below than the critical diameter for full depletion. This sublinearity is characteristic of photoconductors and non-depleted single-nanowire devices. However, planar photoconductors present persistent photoconductivity phenomena and poor UV/visible contrast. The nanowire devices described in this paper present a visible rejection higher than two orders of magnitude and a time response in the hundreds-of-microseconds range, which fits rather the characteristic of single-nanowire devices. The SSL structure displays spectral characteristics and dynamics that depend on bias, and that originate from the presence of opposed internal electric fields along the nanowire, due to the spontaneous polarization in III-nitride materials.

## ◼ ASSOCIATED CONTENT

### Supporting Information

Top-view and cross-section SEM images of the samples under study, HAADF-STEM images of sample SSL, effect of the spinning speed on the planarization process, and effect of the planarization on the sample morphology.

## ◼ ACKNOWLEDGEMENTS

This project received funding from the European Research Council under the European Union's H2020 Research and Innovation programme via the e-See project (Grant No. 758385), and from the French National Research Agency (ANR) via the INMOST project (ANR-19-CE08-0025). A CC-BY public copyright license has been applied by the authors to the present document and will be applied to all subsequent versions up to the Author Accepted Manuscript arising from this submission, in accordance with the grant's open access conditions.

(34) Adivarahan, V.; Simin, G.; Yang, J. W.; Lunev, A.; Khan, M. A.; Pala, N.; Shur, M.; Gaska, R. SiO2-Passivated Lateral-Geometry GaN Transparent Schottky-Barrier Detectors. *Appl. Phys. Lett.* **2000**, *77* (6), 863–865. https://doi.org/10.1063/1.1306647.

(35) Zhang, H.; Guan, N.; Piazza, V.; Kapoor, A.; Bougerol, C.; Julien, F. H.; Babichev, A. V.; Cavassilas, N.; Bescond, M.; Michelini, F.; Foldyna, M.; Gautier, E.; Durand, C.; Eymery, J.; Tchernycheva, M. Comprehensive Analyses of Core–Shell InGaN/GaN Single Nanowire Photodiodes. *J. Phys. D: Appl. Phys.* **2017**, *50* (48), 484001. https://doi.org/10.1088/1361-6463/aa935d.

(36) Jacopin, G.; De Luna Bugallo, A.; Rigutti, L.; Lavenus, P.; Julien, F. H.; Lin, Y.-T.; Tu, L.-W.; Tchernycheva, M. Interplay of the Photovoltaic and Photoconductive Operation Modes in Visible-Blind Photodetectors Based on Axial p-i-n Junction GaN Nanowires. *Appl. Phys. Lett.* **2014**, *104* (2), 023116. https://doi.org/10.1063/1.4860968.

(37) Sanchez-Garcia, M. A.; Calleja, E.; Monroy, E.; Sanchez, F. J.; Calle, F.; Muñoz, E.; Beresford, R. The Effect of the III/V Ratio and Substrate Temperature on the Morphology and Properties of GaN- and AlN-Layers Grown by Molecular Beam Epitaxy on Si(1 1 1). *Journal of Crystal Growth* **1998**, *183* (1–2), 23–30. https://doi.org/10.1016/S0022-0248(97)00386-2.

(38) Yoshizawa, M.; Kikuchi, A.; Mori, M.; Fujita, N.; Kishino, K. Growth of Self-Organized GaN Nanostructures on Al2O3(0001) by RF-Radical Source Molecular Beam Epitaxy. *Jpn. J. Appl. Phys.* **1997**, *36* (4B), L459. https://doi.org/10.1143/JJAP.36.L459.

(39) Ajay, A.; Lim, C. B.; Browne, D. A.; Polaczynski, J.; Bellet-Amalric, E.; den Hertog, M. I.; Monroy, E. Intersubband Absorption in Si- and Ge-Doped GaN/AlN Heterostructures in Self-Assembled Nanowire and 2D Layers. *physica status solidi (b)* **2017**, *254* (8), 1600734. https://doi.org/10.1002/pssb.201600734.

(40) Musolino, M.; Tahraoui, A.; Fernández-Garrido, S.; Brandt, O.; Trampert, A.; Geelhaar, L.; Riechert, H. Compatibility of the Selective Area Growth of GaN Nanowires on AlN-Buffered Si Substrates with the Operation of Light Emitting Diodes. *Nanotechnology* **2015**, *26* (8), 085605. https://doi.org/10.1088/0957-4484/26/8/085605.

(41) Dimkou, I.; Harikumar, A.; Donatini, F.; Lähnemann, J.; den Hertog, M. I.; Bougerol, C.; Bellet-Amalric, E.; Mollard, N.; Ajay, A.; Ledoux, G.; Purcell, S. T.; Monroy, E. Assessment of AlGaN/AlN Superlattices on GaN Nanowires as Active Region of Electron-Pumped Ultraviolet Sources. *Nanotechnology* **2020**, *31* (20), 204001. https://doi.org/10.1088/1361-6528/ab704d.

(42) Consonni, V.; Knelangen, M.; Geelhaar, L.; Trampert, A.; Riechert, H. Nucleation Mechanisms of Epitaxial GaN Nanowires: Origin of Their Self-Induced Formation and Initial Radius. *Phys. Rev. B* **2010**, *81* (8), 085310. https://doi.org/10.1103/PhysRevB.81.085310.

(43) Hille, P.; Müßener, J.; Becker, P.; de la Mata, M.; Rosemann, N.; Magén, C.; Arbiol, J.; Teubert, J.; Chatterjee, S.; Schörmann, J.; Eickhoff, M. Screening of the Quantum-Confined Stark Effect in AlN/GaN Nanowire Superlattices by Germanium Doping. *Applied Physics Letters* **2014**, *104* (10), 102104. https://doi.org/10.1063/1.4868411.

(44) Ajay, A.; Lim, C. B.; Browne, D. A.; Polaczyński, J.; Bellet-Amalric, E.; Bleuse, J.; den Hertog, M. I.; Monroy, E. Effect of Doping on the Intersubband Absorption in Si- and Ge-Doped GaN/AlN Heterostructures. *Nanotechnology* **2017**, *28* (40), 405204. https://doi.org/10.1088/1361-6528/aa8504.

(45) Sabelfeld, K. K.; Kaganer, V. M.; Limbach, F.; Dogan, P.; Brandt, O.; Geelhaar, L.; Riechert, H. Height Self-Equilibration during the Growth of Dense Nanowire Ensembles: Order Emerging from Disorder. *Appl. Phys. Lett.* **2013**, *103* (13), 133105. https://doi.org/10.1063/1.4822110.

(46) Songmuang, R.; Ben, T.; Daudin, B.; González, D.; Monroy, E. Identification of III–N Nanowire Growth Kinetics via a Marker Technique. *Nanotechnology* **2010**, *21* (29), 295605. https://doi.org/10.1088/0957-4484/21/29/295605.

(47) Guerfi, Y.; Doucet, J. B.; Larrieu, G. Thin-Dielectric-Layer Engineering for 3D Nanostructure Integration Using an Innovative Planarization Approach. *Nanotechnology* **2015**, *26* (42), 425302. https://doi.org/10.1088/0957-4484/26/42/425302.
19

**Figures**

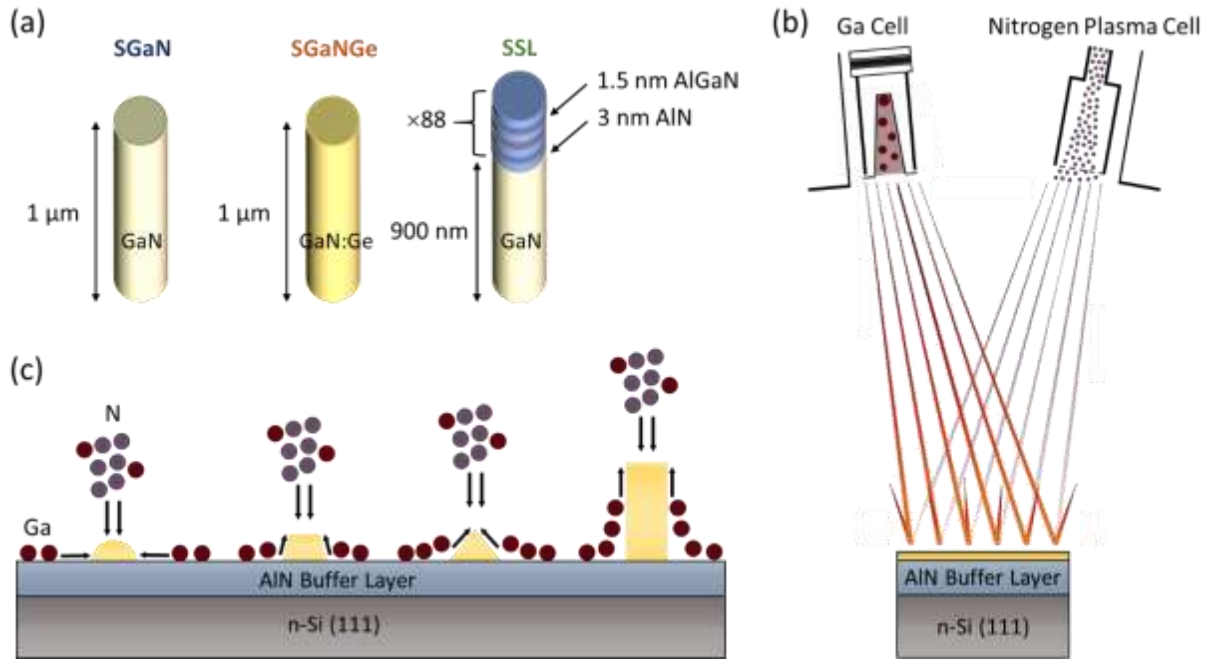

**Figure 1.** Schematic representation of (a) samples under study, nanowire growth process (b) MBE structure, and (c) nanowire nucleation mechanism (Adapted from ref [42]).

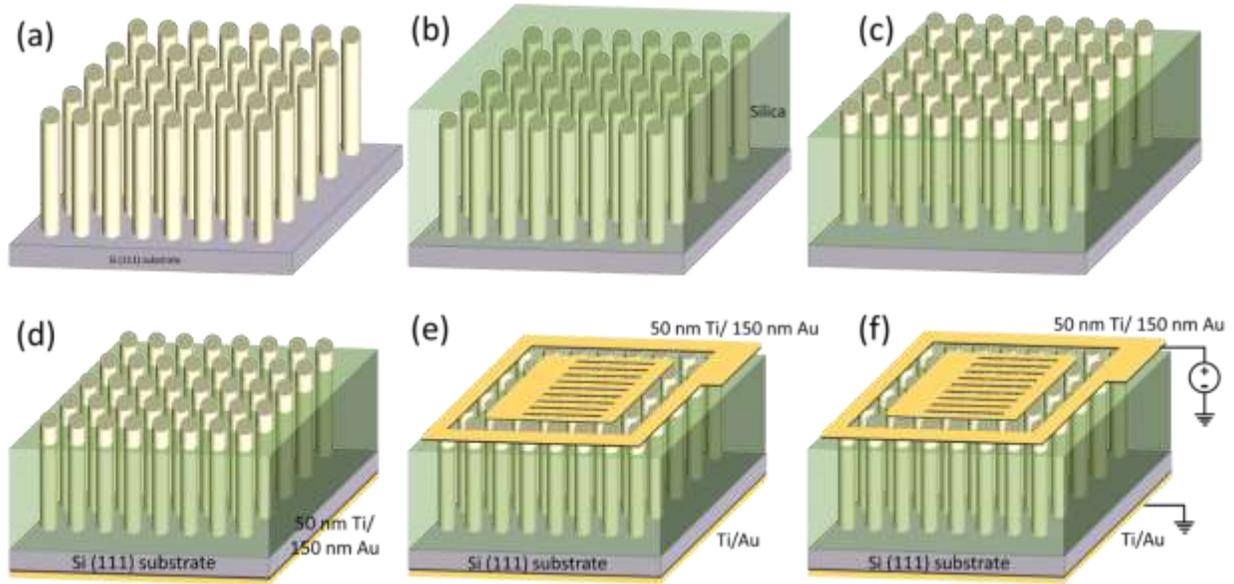

**Figure 2.** Schematic representation of the device fabrication steps: (a) As-grown nanowire sample, (b) planarization, (c) exposure of the nanowire tip after dry etching, (d) back contact on the silicon substrate, (e) first lithography and metallization to define the fingers and contact pad, and (f) second lithography and metallization to deposit a semi-transparent spread layer.

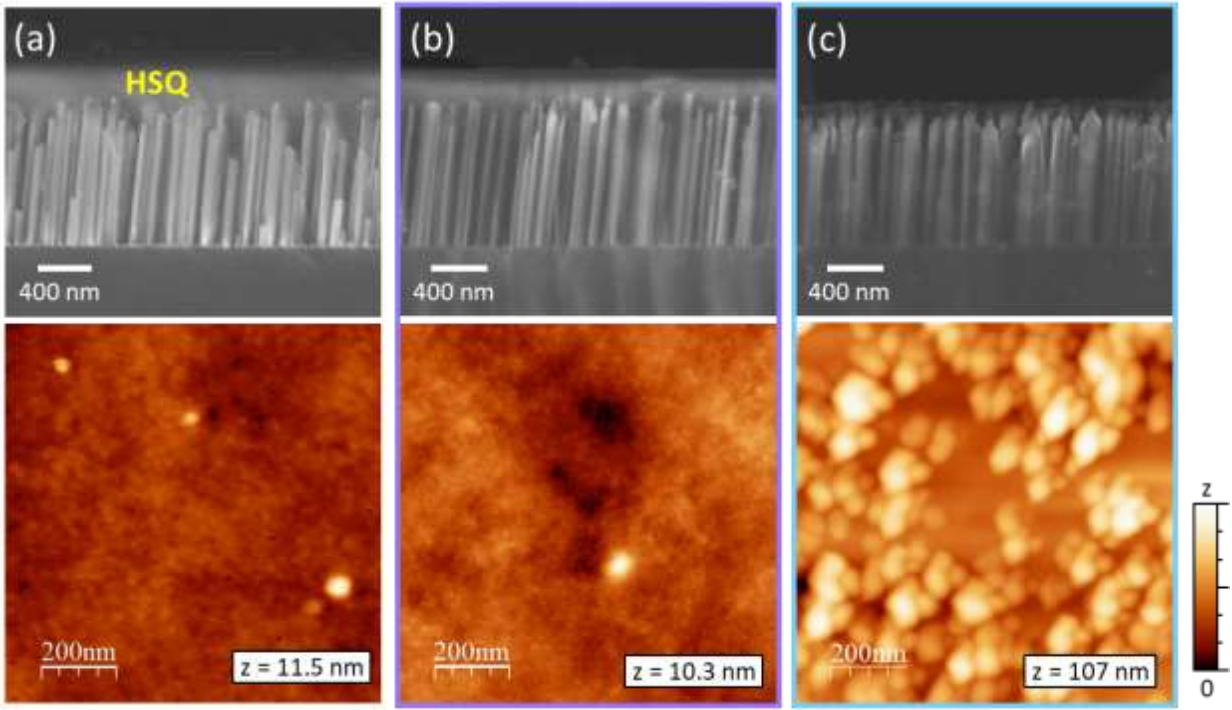

**Figure 3.** (Top) Cross-section SEM image and (bottom) AFM image of sample SGaN after planarization and (a) before $SiO_2$ etching, (b) after etching 150 nm of $SiO_2$ and (c) after etching 350 nm of $SiO_2$.

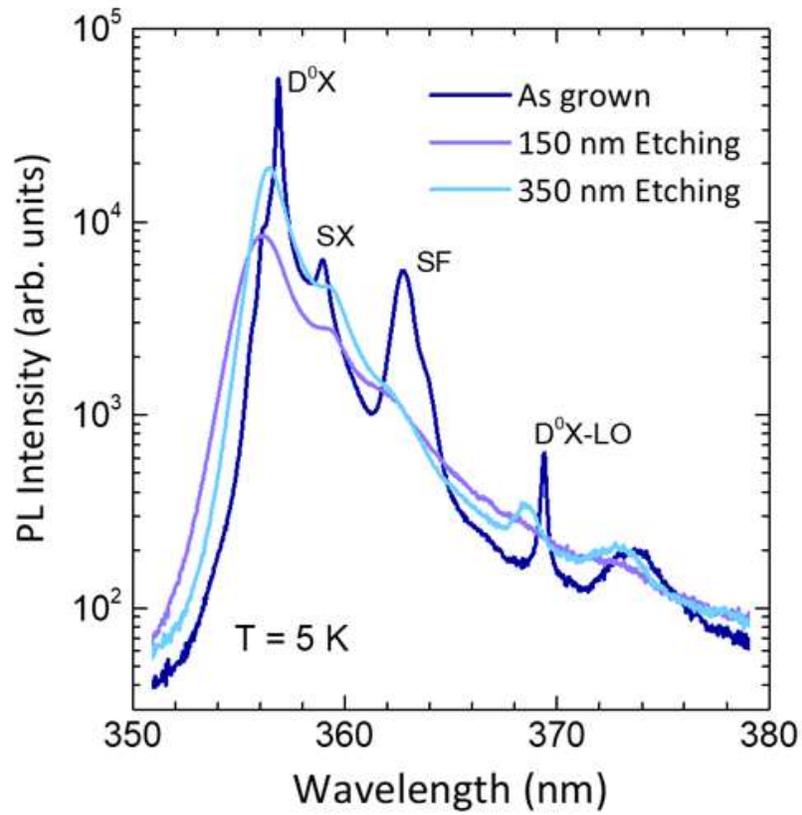

**Figure 4.** Low temperature (T = 5 K) PL spectrum of as-grown, 150-nm-etched and 350-nm-etched SGaN.

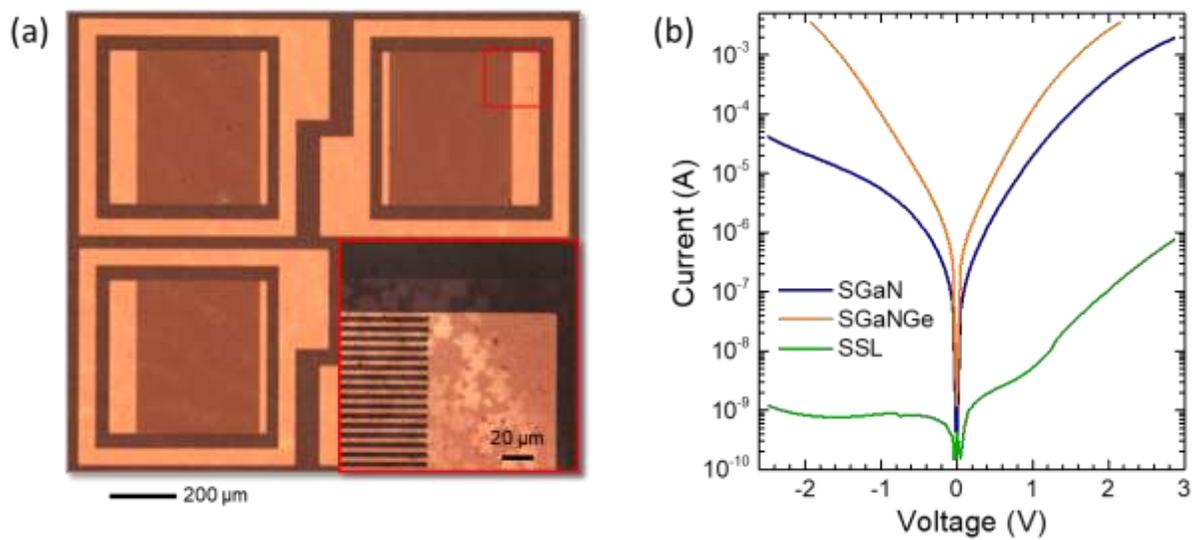

**Figure 5.** (a) Top view of the devices in an optical microscope, with a zoom on the fingers. (b) Current-voltage characteristics of samples SGaN, SGaNGe and SSL measured in the dark.

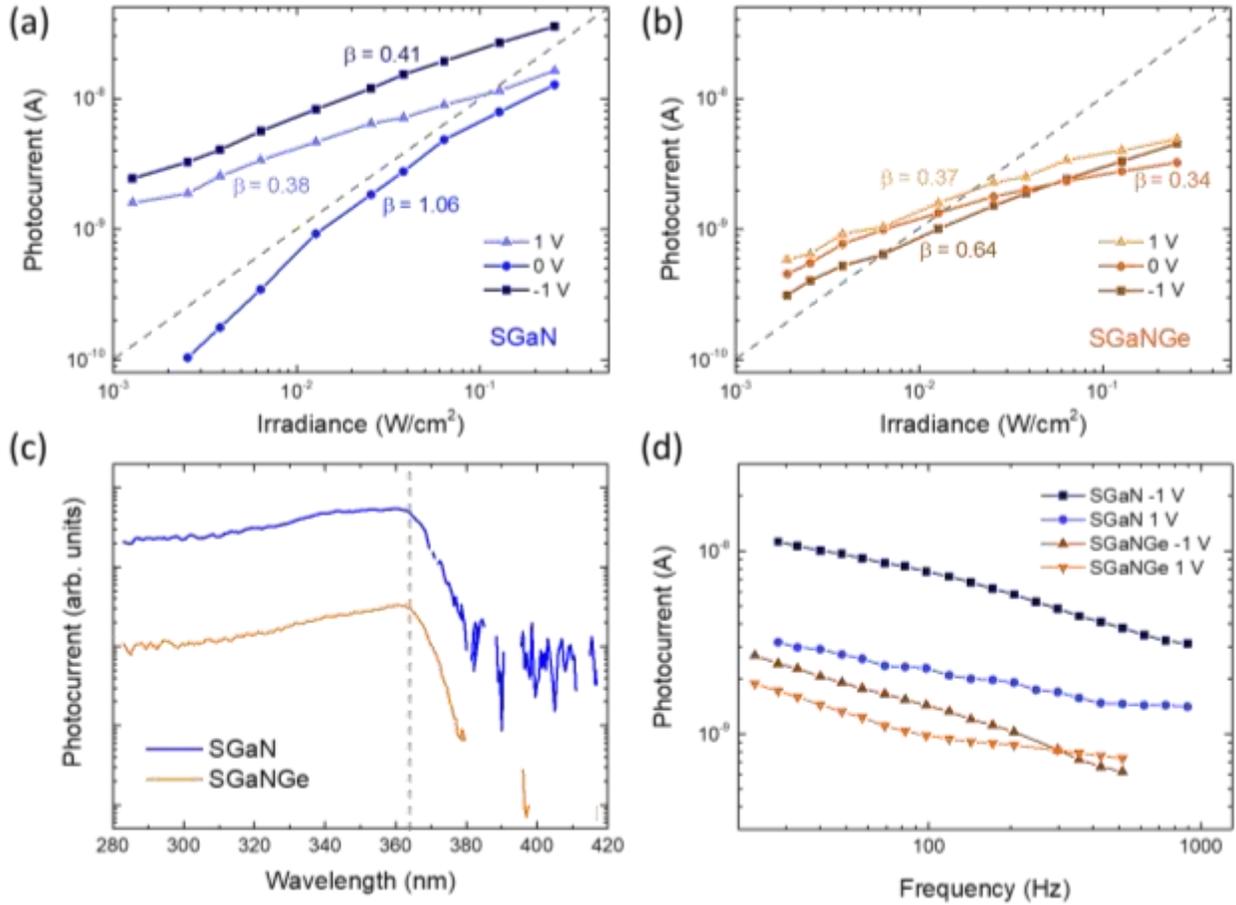

**Figure 6.** Variation of photocurrent ($I_{ph}$) as a function of impinging optical power ($P_{opt}$) at 325 nm for (a) SGaN, (b) SGaNGe, comparison between SGaN and SGaNGe samples in terms of (c) Spectral response measurement, and (d) Variation of the photocurrent as a function of the chopping frequency (SGaN and SGaNGe are indicated with blue and orange color, respectively).

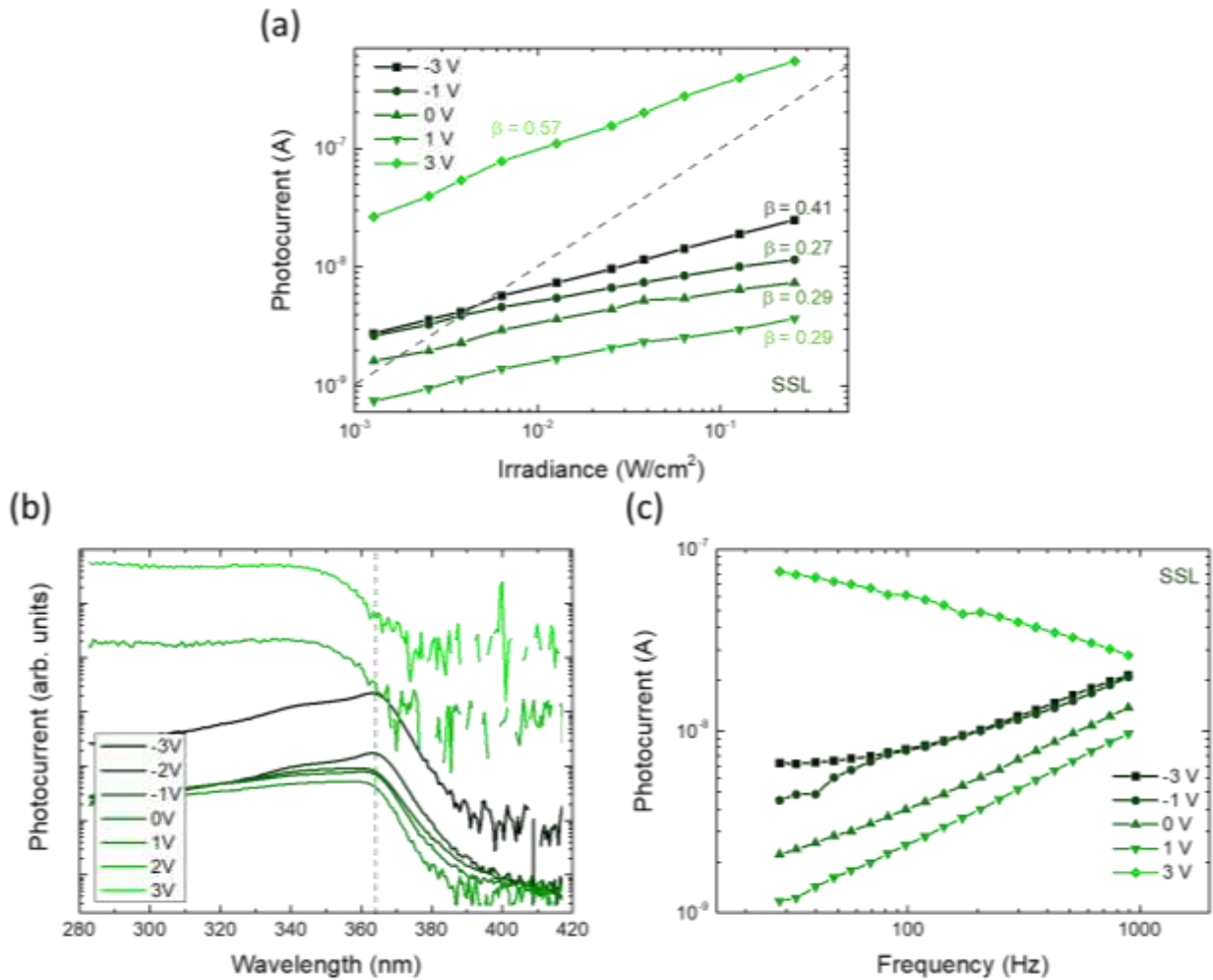

**Figure 7.** Optical measurements of sample SSL (a) Variation of photocurrent ($I_{ph}$) as a function of impinging optical power ($P_{opt}$) at 325 nm (The slope for $\beta=1$ (i.e. linear behavior) is represented by a gray dashed line and the vertical gray line indicates the power the measurements are taken), (b) Spectral response (the gray dashed line indicates the wavelength of the GaN at room temperature), and (c) Variation of the photocurrent as a function of the chopping frequency.

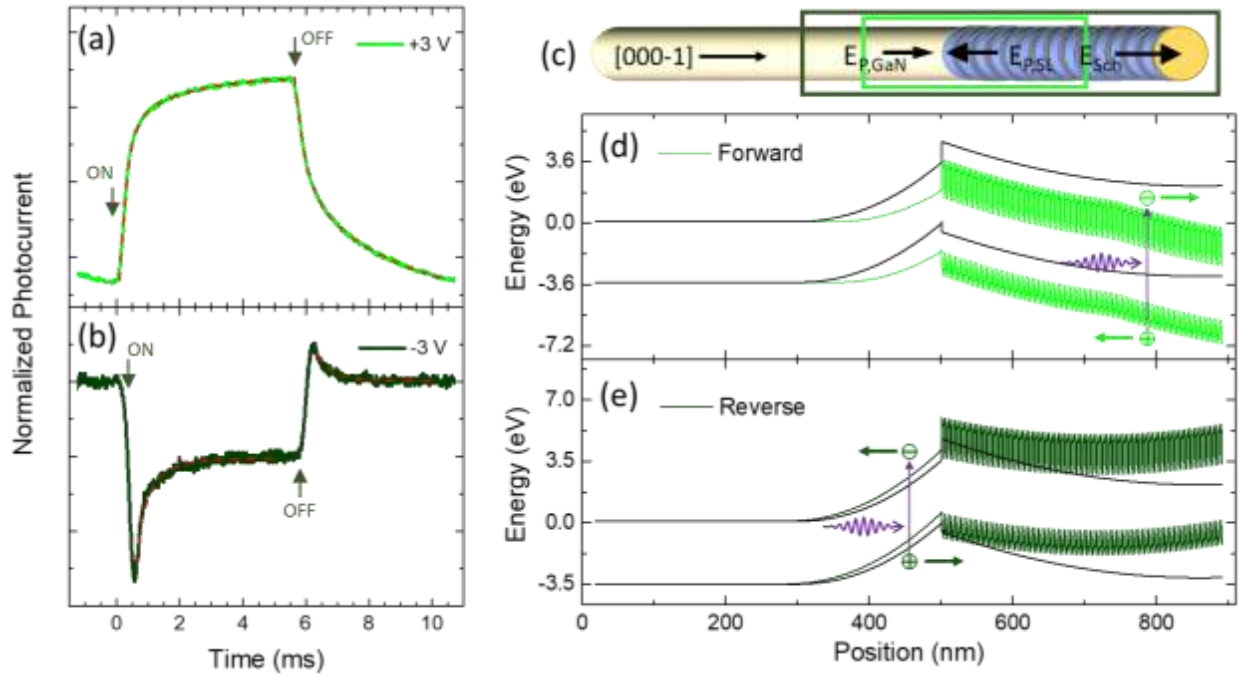

**Figure 8.** Rise and fall time of the photocurrent measured in the SSL sample at (a) +3 V and (b) -3 V. Dashed lines are biexponential fits. (c) Schematic representation of the SSL structure given with the internal electric field formations within the structure. (d-e) One-dimensional calculations of the SSL band structure along the growth axis under (d) forward and (e) reverse bias. The black band diagram corresponds to the potential profile of the nanowire structure under zero bias (the quantum wells were smoothed out in this reference profile).

# Ultraviolet Photodetectors based on GaN and AlGaN/AlN Nanowire Ensembles: Effects of Planarization with Hydrogen Silsesquioxane and Nanowire Architecture

E. Akar[†,*], I. Dimkou[‡], A. Ajay[†], Martien I den Hertog[§], and E. Monroy[†]

[†] Univ. Grenoble-Alpes, CEA, Grenoble INP, IRIG, PHELIQS, 17 av. des Martyrs, 38000 Grenoble, France
[‡] Univ. Grenoble-Alpes, CEA, LETI, 38000 Grenoble, France
[§] Univ. Grenoble-Alpes, Institut Néel-CNRS, 25 av. des Martyrs, 38000 Grenoble, France

*Corresponding author: elcin.akar@cea.fr

## Samples under study

Top-view and cross-section SEM images of the samples under study are shown in Figure S1 along with schematic representation of their structures.

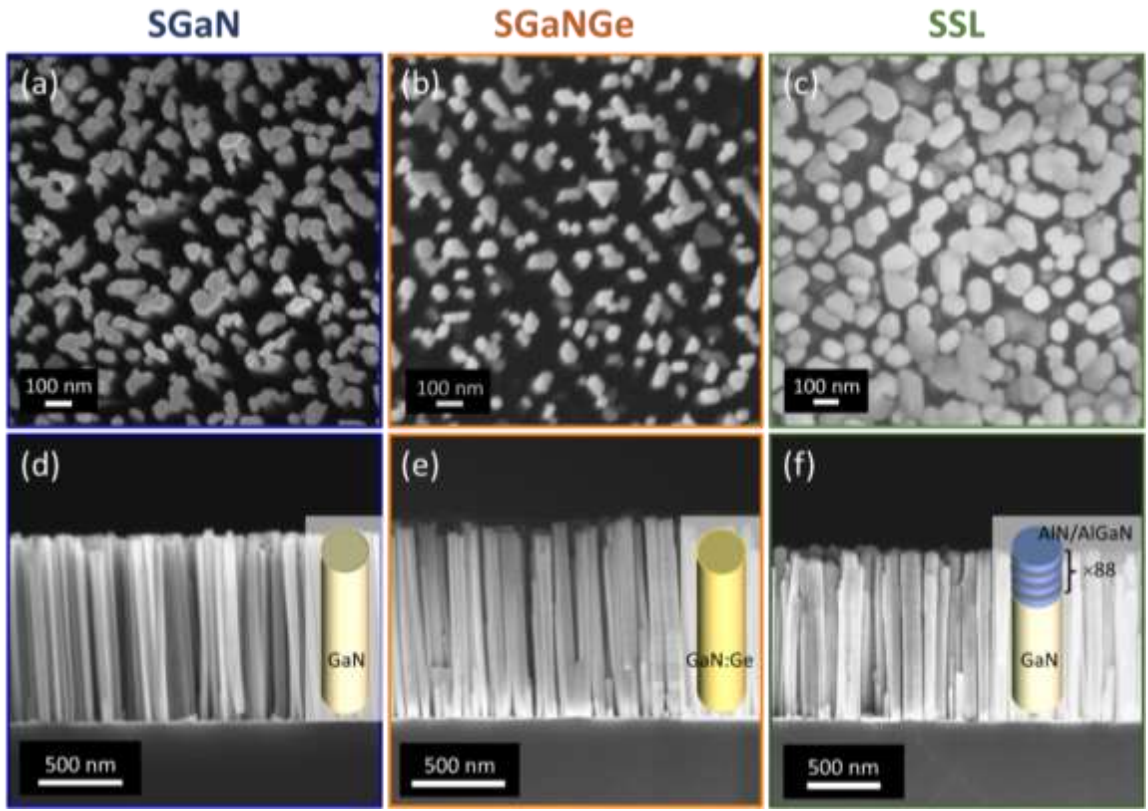

**Figure S9.** (a-c) Top-view and (d-f) cross-section SEM images of the samples under study: (a,d) SGaN, (b,e) SGaNGe and (c,f) SSL. The schematic description of one nanowire appears as an inset in images (d-f).

Depending on the structural configuration and the growth conditions, the samples may present a high degree of coalescence. For instance, in the case of SSL the growth of an AlGaN/AlN superlattice structure on top of the GaN nanowires leads to an increment in coalescence of the nanowires.

## HAADF-STEM images of SSL

Figure S2 presents high-angle annular dark-field scanning transmission electron microscopy (HAADF-STEM) micrographs of the SSL sample with an AlGaN/AlN superlattice. The sample was prepared by mechanical dispersion on a holey carbon grid for observation. For this reason, mostly bundles of nanowires are observed. Higher magnification of a single nanowire exhibits an abrupt interface between GaN stem and superlattice. From the active region, the AlGaN quantum well and AlN barrier thickness are extracted as 1.5 nm and 3 nm, respectively which is in accordance with the experimental growth parameters of sample.

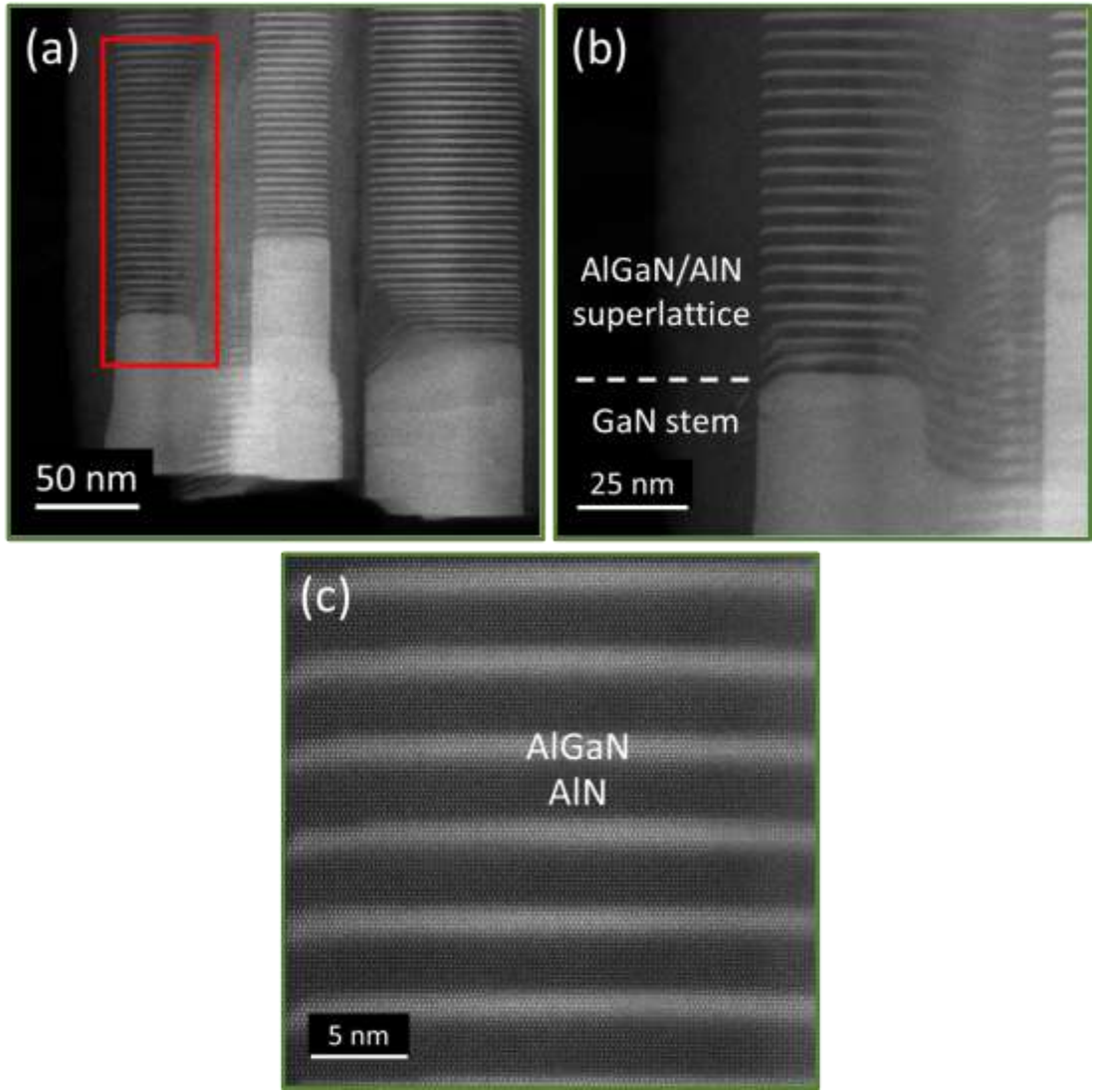

**Figure S10.** HAADF-STEM images of sample SSL (a) bundle of nanowires, (b) higher magnification micrograph on the region indicated in (a) on a single nanowire and (c) its active region oriented on a [2-1-10] zone axis, respectively.

## Effect of the spinning speed on the planarization process

To planarize the nanowire ensembles, hydrogen silsesquioxane (HSQ) is deposited using the spin coating technique. The coating quality in terms of penetration and wetting does not vary as a function of the spinning speed, yet the thickness of the coating decreases for faster rotation, as shown in Table S1. Note the difference between the total thickness of the HSQ-coated sample and the nominal HSQ thickness on a flat substrate. This means that the penetration of HSQ in the nanowire ensemble delays the spreading, resulting in HSQ layers significantly thicker than the nominal thickness. This also implies that the total thickness should depend on the nanowire shape and density. Therefore, deviations from sample to sample are expected.

**Table S1.** Total thickness of HSQ-coated SGaN as a function of the spinning speed.

| Spinning speed | HSQ-coated thickness | Nominal HSQ Thickness |
|---|---|---|
| 1000 rpm | 1.65 µm | > 0.81 µm |
| 2000 rpm | 1.41 µm | 0.81 µm |
| 3000 rpm | 1.34 µm | 0.69 µm |
| 4000 rpm | 1.25 µm | 0.58 µm |
| 5000 rpm | 1.39 µm | < 0.58 µm |

## Effect of the planarization on the sample morphology

The SEM images of sample SGaN before and after planarization at 2000 rpm are given in Figure S3a and b. It is clear from the image that HSQ fills the gap between nanowires after the spin coating. As discussed before, the planarization quality does not vary with spin-coating speed, but it may change from sample to sample due to structural alterations. Some samples such as SSL exhibit coalescence due to the lateral growth of the nanowires during the deposition of the AlGaN/AlN superlattice (see Figure S1). Hence, gap filling with HSQ is more difficult for such sample which may cause reduced wetting behavior along with small voids in the structure, outlined with arrows in the Figure S3c.

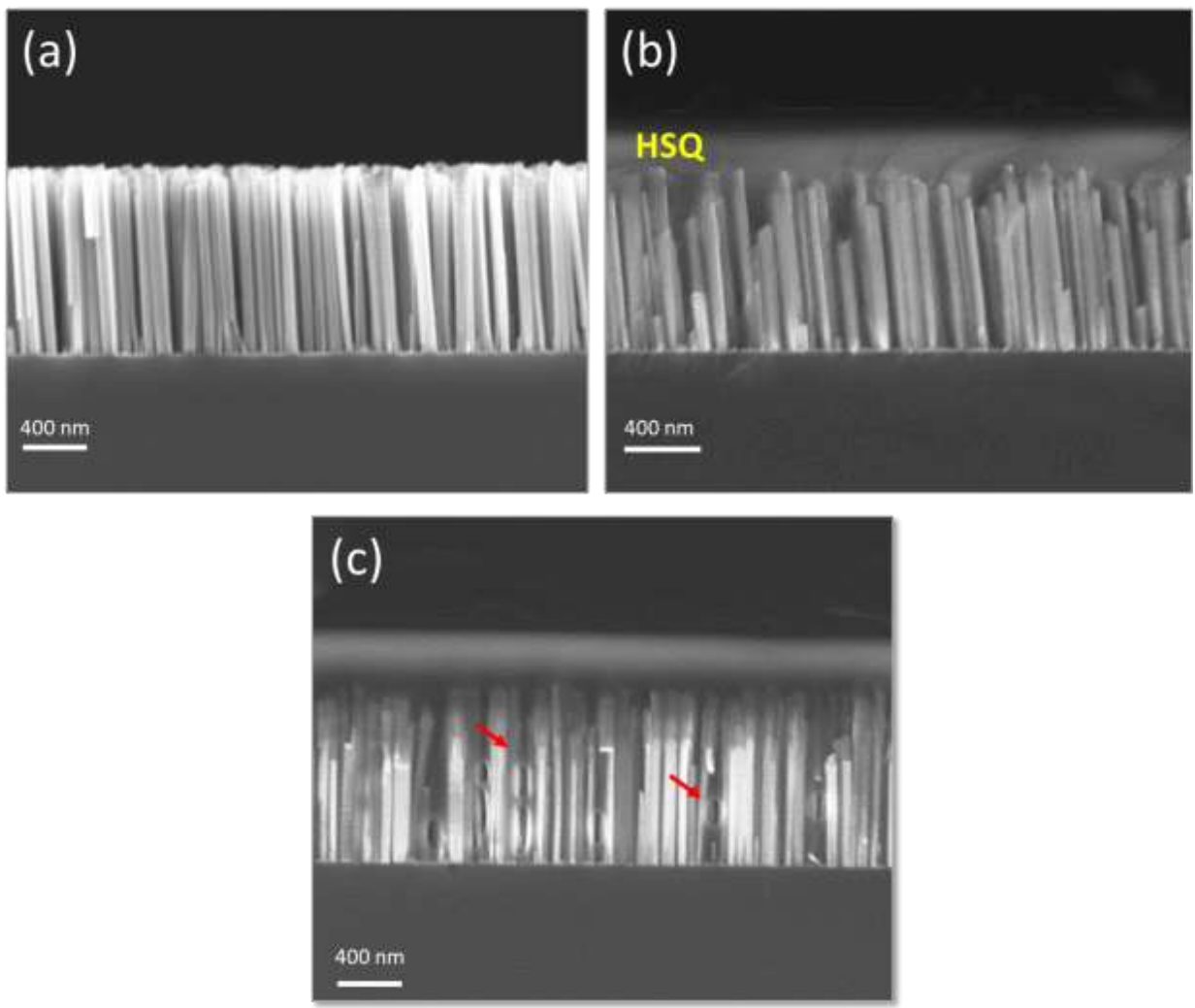

**Figure S3**. SEM images of SGaN (a) before and (b) after the planarization at 2000 rpm, and (c) SSL after planarization at 5000 rpm. Red arrows indicate small voids within the sample.